\documentclass[twocolumn, prl, 10pt,superscriptaddress]{revtex4-1}
\usepackage{graphicx}
\usepackage{amsmath}
\usepackage{longtable}
\usepackage{latexsym}
\usepackage{times}
\usepackage{graphics}
\usepackage{graphicx}
\usepackage{epsfig}
\usepackage{wrapfig}
\usepackage[T1]{fontenc}
\usepackage{float}
\usepackage{enumerate}
\usepackage{setspace,ifthen}
\usepackage{amsthm} 
\usepackage{amsmath}
\usepackage{amssymb}	
\usepackage[usenames,dvipsnames]{xcolor}
\usepackage{mathrsfs,amsmath}
\usepackage{mathrsfs}
\usepackage{units}
\usepackage{braket}

\def\XXint#1#2#3{{\setbox0=\hbox{$#1{#2#3}{\int}$}
     \vcenter{\hbox{$#2#3$}}\kern-.5\wd0}}

\begin{document}

\title{Application of optimal band-limited control protocols to quantum noise sensing}
\date{\today}

\author{V. M. Frey}
\affiliation{ARC Centre for Engineered Quantum Systems, School of Physics, The University of Sydney, NSW 2006 Australia, and \\
National Measurement Institute, West Lindfield NSW 2070 Australia}

\author{S. Mavadia}
\affiliation{ARC Centre for Engineered Quantum Systems, School of Physics, The University of Sydney, NSW 2006 Australia, and \\
National Measurement Institute, West Lindfield NSW 2070 Australia}

\author{L. M. Norris}
\affiliation{\mbox{Department of Physics and Astronomy, Dartmouth 
College, 6127 Wilder Laboratory, Hanover, NH 03755, USA}}

\author{\mbox{W. de Ferranti}}
\affiliation{ARC Centre for Engineered Quantum Systems, School of Physics, The University of Sydney, NSW 2006 Australia, and \\
National Measurement Institute, West Lindfield NSW 2070 Australia}

\author{D. Lucarelli} 
\affiliation{\mbox{Johns Hopkins University, Applied Physics Laboratory,
11100 Johns Hopkins Road, Laurel, MD 20723, USA}}

\author{L. Viola}
\affiliation{\mbox{Department of Physics and Astronomy, Dartmouth 
College, 6127 Wilder Laboratory, Hanover, NH 03755, USA}}

\author{M. J. Biercuk$ ^{\dagger} $}
\affiliation{ARC Centre for Engineered Quantum Systems, School of Physics, The University of Sydney, NSW 2006 Australia, and \\
National Measurement Institute, West Lindfield NSW 2070 Australia}
\email{\emph{These two authors contributed equally to this work} \newline $^{\dagger}$Contact: michael.biercuk@sydney.edu.au }

\maketitle
\textbf{Industrial, metrological, and medical applications provide a strong technological pull for advanced nanoscale sensing techniques~\cite{schulte2005nanotech}.  An emerging area of interest relates to accessing and exploiting the exquisite sensitivity of quantum coherent systems to their surrounding environment as a means to augment sensor performance~\cite{HollenBerg_Sensing, Ajoy17, maze2008nanoscale}.  
 Essential to the functionality of qubit-based sensors is the availability of control protocols which shape their 
response in frequency space.  However, a major challenge is that common control routines result in out-of-band spectral leakage which complicates interpretation of the sensor's signal~\cite{degen_retract, degen_followup}.  In this work we demonstrate provably optimal narrowband controls ideally suited to spectral estimation of a qubit's noisy environment.  Experiments with trapped ions demonstrate reduction of spectral leakage by orders of magnitude over conventional controls when a near resonant driving field is modulated by discrete prolate spheroidal sequences (aka Slepian functions~\cite{Slepian1978}).  We tune the narrowband sensitivity using concepts from RF engineering and experimentally reconstruct complex noise spectra.  We then deploy these techniques to identify previously immeasurable frequency-resolved amplitude noise in our qubit's microwave synthesis chain with calibrated sensitivity better than 0.001 dB.}

\begin{figure}[bp]
\includegraphics[scale=1]{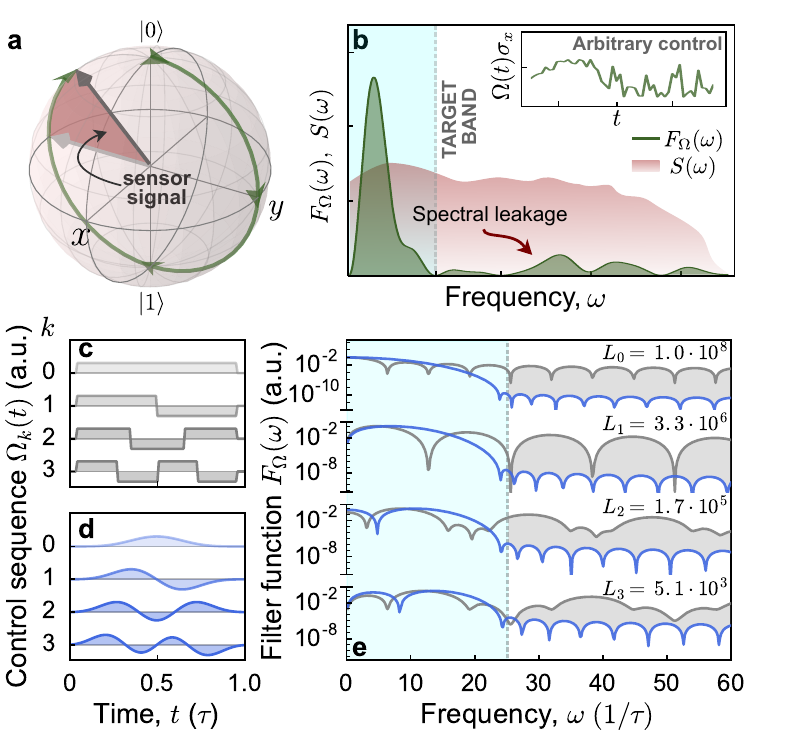}
\caption{Schematic illustration of frequency responses of time-domain control protocols. \textbf{a}) Sensor signal for a driven qubit is derived from the spectral overlap of the noise and control, producing a detectable rotation error. \textbf{b}) Sensor spectral response for a given control protocol (inset) may show leakage outside of the desired target band, which makes an undesired contribution to the signal in panel \textbf{a}.   \textbf{c-d}). Sample control sequences in the time-domain for flat-top echo protocols and their DPSS-modulated complements. Orders shown here: $k=0,1,2,3$, and $NW=4$.
\textbf{e}) $F_{\Omega}(\omega)$ capturing the frequency response of the control envelopes in \textbf{c,d}. The dotted line indicates the boundary of the target band along the positive frequency axis, 
$[0, \omega_B]$, $\omega_B = 2\pi W/\Delta t$, where $\tau=N\Delta t$ is the duration of the control.  Inferior spectral concentration manifested as spectral weight of flat-top pulses outside of the target band is highlighted with shading. Relative power outside of the target band comparing $k$th-order DPSS ($\mathrm{D}$) and flat-top ($\mathrm{FT}$) pulses is calculated as $L_{k} = (1 - \lambda_{\mathrm{FT}}) / (1 - \lambda_{\mathrm{D}})$, where 
$\lambda_\ell \equiv  \int_0^{\omega_\mathrm{B}} \mathrm{d} \omega F_{\Omega}^{(\ell)}(\omega) / \int_0^{\infty} \mathrm{d} \omega F_{\Omega}^{(\ell)}(\omega)$.}
\label{fig1}
\end{figure}

Qubits naturally exhibit broadband coupling to their environments, but the application of a temporally modulated driving field 
can alter their frequency response in a desired way.  For instance, application of modulation which periodically flips the qubit's state has allowed for a narrowband spectral response~\cite{Biercuk_Filter}, which may be tuned by adjusting the interpulse spacing or extending the sequence duration.  This general approach to ``dynamical decoupling noise spectroscopy'' has seen broad adoption in quantum information~\cite{Alvarez_Spectroscopy,Bylander2011, Ozeri2013, Morello2014, Norris_Spectroscopy,Paz2017}, as well as in nanoscale diamond sensors for biomedical and physics-based applications~\cite{ajoy2015atomic, le2013optical, say2011luminescent, Hirose12}.  
However, control implemented in this form suffers from the significant complication of spectral leakage, where 
signals outside of a target sensing frequency band can contribute to the sensor's response (Fig.~\ref{fig1}b), and if not 
properly accounted for, can lead to bias when estimating the spectral density of a signal from experimental data ~\cite{degen_followup}. 

In an ideal case, for frequency-domain spectral-estimation applications, the chosen control protocol applied to the qubit would be sensitive only within a user-determined band. Pulsed dynamical decoupling is often employed because the leading component of the filter transfer function describing the modulated sensor's performance is narrowly peaked~\cite{Biercuk_Filter, Bylander2011}.  Examination of the so-called control propagator describing the time-domain response of a qubit subject to this control, however, reveals that the effective square-wave form of the control propagator (Fig.~\ref{fig1}c) leads to the appearance of an infinite chain of harmonics in the Fourier domain (Fig.~\ref{fig1}d).  These out-of-band harmonics can then contribute bias in noise spectroscopy protocols.

The problem of spectral leakage is well known in classical signal processing
and has led to the development of time-bandwidth optimized functions, which we identify here as useful in quantum sensing applications.  Instead of performing modulation on the qubit sensor which results in a flat-top response (\emph{e.g.}, pulsed dynamical decoupling or instantaneous phase-flips under driven rotary-echo~\cite{Gust2012Driven, Ball2014}), we employ continuously modulated pulse envelopes described by {\em discrete prolate spheroidal sequences} (DPSS) \cite{Slepian1978}. These are an orthogonal set of discrete-time functions that minimize the energy outside a predefined frequency band. The DPSS have in the past found application in time-domain signal processing~\cite{thompson_multitaper}, and are now widely employed across classical spectral analysis and estimation \cite{percival1993spectral, red_book}. Additionally, DPSS have been suggested in magnetic resonance imaging to avoid out of band excitation, known as the Gibbs artefact \cite{Hasenfeld1985}. This strong base of demonstrations motivates our use of DPSS in the quantum sensing setting.

For a time-domain sequence consisting of $N$ elements, 
characterized by sampling interval $\Delta t$, and half-bandwidth parameter $W\in (0,1/2]$, 
the DPSS may be defined as real solutions to the eigenvalue problem 
\begin{equation*}
\sum_{m=0}^{N-1} \!\frac{\sin 2\pi W(n-m)}{\pi(n-m)} \, v_m^{(k)} (N,W) = \lambda_k(N, W) v_n^{(k)}(N, W), 
\end{equation*}
where $v_m^{(k)}(N, W)$ is the $m$th element of the $k$th-order DPSS for $k, m \in  \{ 0, 1, \ldots , N -1\}$.  The discrete Fourier transform of $v_m^{(k)}(N, W)$ into the (angular) frequency domain $[-\pi/\Delta t,\pi /\Delta t]$ is the discrete prolate spheroidal wavefunction, $U^{(k)}(N, W; \omega)$, which is spectrally concentrated in the target band $[-\omega_B, \omega_B]\equiv [-2\pi W\!/\Delta t,2\pi W\!/\Delta t]$. The eigenvalue $\lambda_k(N,  W)$ directly quantifies the spectral concentration of $U^{(k)}(N, W; \omega)$, 
\emph{i.e.}, 
the fraction of spectral power within the target band compared to the total spectral power, as 
$\lambda_k(N,  W)  = \frac{\int_{-\omega_B}^{\omega_B} U^{(k)}(N, W; \omega)^2}
{\int_{-\pi/\Delta t}^{\pi/\Delta t} U^{(k)} (N, W; \omega)^2}.$ 
For a fixed choice of $N$ and $W$, the spectral concentration is maximized for the lowest-order $k=0$, 
and decreases with increasing $k$.  For example, the DPSS of order $k<2 \lfloor NW\rfloor -1$ have $\unit[\ge 70]{\%}$ of their spectral weight within the target band (Fig.~\ref{fig1}d).  

\begin{figure}[tp!]
\includegraphics[scale=1]{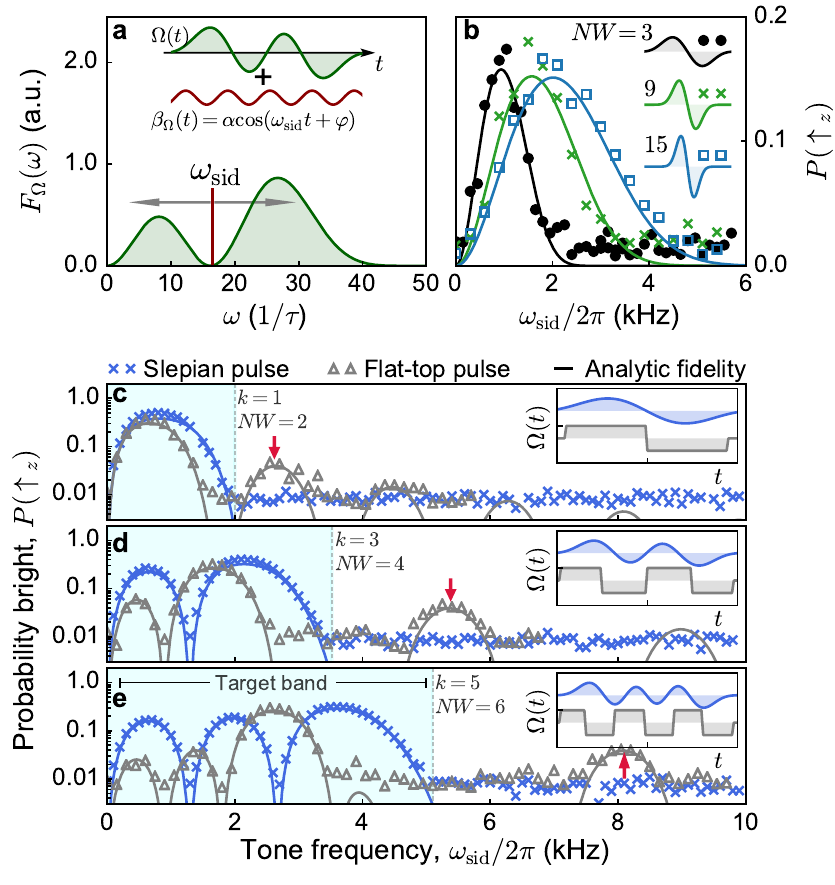}
\caption{Spectral reconstruction of control filter transfer functions. \textbf{a)} A single-frequency modulation, $\beta_{\Omega}(t)$ is 
applied to the control envelope $\Omega(t)$ which drives rotations about ${x}$. Measuring operational fidelity for each $\omega_{\mathrm{sid}}$
allows frequency-resolved reconstruction of $F_{\Omega}(\omega)$.  \textbf{b)} Experimental reconstruction of DPSS filter
functions for $k=1$ and varying $NW$. Control envelopes (shown schematically as insets) have a duration of $\unit[1.1]{ms}$ with area normalised to $\pi$ prior to modulation, and measurements are averaged over 10 linearly sampled $\varphi\in[0,2\pi]$, with $\alpha=0.5$. Each phase realisation is repeated 50 times to reduce the influence of photon shot noise.  \textbf{c-e)} Comparison of the measured spectral response $F_{\Omega}(\omega)$ of flat-top and DPSS control sequences implementing 
${\mathbb{I}}$ for different $k$ and $NW$ on a semilog plot.  Shading indicates the target band over which spectral response should be concentrated. We choose $NW=k+1$ for each order $k$, to conservatively maintain spectral concentration of the DPSS while matching the number of zero crossings in the flat-top protocols.  Markers represent experimental measurements and solid lines show the analytic $\mathcal{F}_{\mathrm{av}}$ calculated based on $F_{\Omega}(\omega)$. Arrows highlight out-of-band sensitivity due to harmonics of the flat-top control. We employed large modulation depths ($\alpha=0.95$ for DPSS, $\alpha=0.85$ for flat-top pulses) to amplify these signals.  Measurement sensitivity floor $\sim 0.5\%$.
The relative duration of the pulses is fixed, resulting in slightly broader bandwidths within the target band for the DPSS pulses relative to flat-top.} 
\label{fig2}
\end{figure}

To characterize the frequency response of a qubit-based sensor undergoing an arbitrary control protocol in the presence of universal, 
multi-axis classical noise, we rely on the filter-transfer function formalism~\cite{Kofman2001,Biercuk_Filter, SoareNatPhys2014, GreenPRL2012, GreenNJP2013,PazFFF}. The qubit sensor's response to its environment under the application of control is given approximately by the measured fidelity of the operation.  This is captured, in the weak-noise limit, as the spectral overlap of the noise power spectral density in multiple quadratures, $S_{i}(\omega)$, with a transfer function describing the control, $F_{i}(\omega)$, as \mbox{$1-\mathcal{F}_{\mathrm{av}} \approx\exp[{-\pi^{-1} \sum_{i=\Omega, z} \int \mathrm{d}\omega F_i(\omega) S_i(\omega)}]$}.  
Here, the sum is taken over amplitude noise contributions proportional to the applied control, \emph{i.e.}, the qubit Rabi rate 
$\propto \Omega \sigma_x$, and dephasing components $\propto \sigma_z$.  The presence of a signal in the sensor's target band, defined by the applied modulation, will be manifested as a reduction in the fidelity of the operation implemented (here we rely exclusively on projective measurement onto the qubit basis states).  The target operation may in general be either the identity or another nontrivial quantum state transformation (Fig.~\ref{fig1}a).  

Analytic calculation of $F_{\Omega}(\omega)$ both for flat-top modulation (commonly associated with dynamical decoupling protocols and here a rotary spin echo) and for piecewise-constant modulation defined by the DPSS, reveals the superior spectral concentration of the latter (Fig.~\ref{fig1}e).  
While the main lobe of $F_{\Omega}(\omega)$ is broader inside the target band (blue shading) for DPSS modulation as compared to the rotary spin echo, leakage outside the target band is significantly suppressed. For the rotary spin echo, spectral leakage increases out-of-band sensitivity by $30-80$ dB relative to the DPSS, indicated by the value $L_{k}$ (Fig.~\ref{fig1}e).   

We perform experiments to directly test the spectral response of a driven qubit-based sensor using trapped \mbox{$^{171}$Yb$^{+}$ ions}, where the qubit is realized through the hyperfine splitting of the $^1$S$_{1/2}$ ground state with a transition frequency $\sim\unit[12.6]{GHz}$. We can modulate the amplitude and phase of the driving microwave field arbitrarily using a vector signal generator, providing full control of the qubit state on the Bloch sphere.  We employ projective measurements of the qubit state in the $z$-basis and average over experiments to identify deviations from ideal control operations which constitute our signal of interest, $P(\uparrow_z)$. 
Details of the experimental system appear in~\cite{Soare2014bath, MavadiaNatComms2017} and in the \emph{Supplementary Materials}.

We verify the spectral properties of DPSS-controlled qubit sensors by performing frequency-selective system identification \cite{stengel2012optimal} to map out the effective spectral response of the driven qubit (Fig.~\ref{fig2}a).  A small single-frequency modulation, \mbox{$\beta_{\Omega}(t) = \alpha \cos(\omega_{\mathrm{sid}}t  + \varphi)$} is added to the applied control envelope of the driving field, 
producing \mbox{$\Omega(t) \mapsto \Omega(t)(1 + \beta_{\Omega}(t))$}.  By scanning the tunable modulation frequency $\omega_{\mathrm{sid}}$ and averaging over phase $\varphi$ for fixed modulation depth $\alpha$, we effectively reconstruct the filter transfer function of the control, $F_{\Omega}(\omega)$, captured in the average fidelity metric $\mathcal{F}_{\mathrm{av}}$ for the applied operation.
Experimental reconstruction of the qubits' spectral response under DPSS modulation for $k=1$ shows good agreement with the analytically calculated fidelity, in addition to the expected broadening as $NW$ is increased (Fig.~\ref{fig2}b).  Using the same technique, we can experimentally compare the frequency response of qubits driven by pulses with DPSS envelopes to their flat-top counterparts, 
as shown in Fig.~\ref{fig2}c-e.  These experiments demonstrate the superior spectral concentration in the target band (shaded region) defined by the product $NW$ produced by DPSS modulation. Measurements on DPSS-modulated qubits show no detectable sensitivity to perturbations 
(given by $\beta_{\Omega}(t)$) outside of the target band, 
but the flat-top modulated experiments exhibit harmonics (marked by arrows), constituting a source of spectral leakage in sensing applications.

\begin{figure}[t!]
\includegraphics[scale=1]{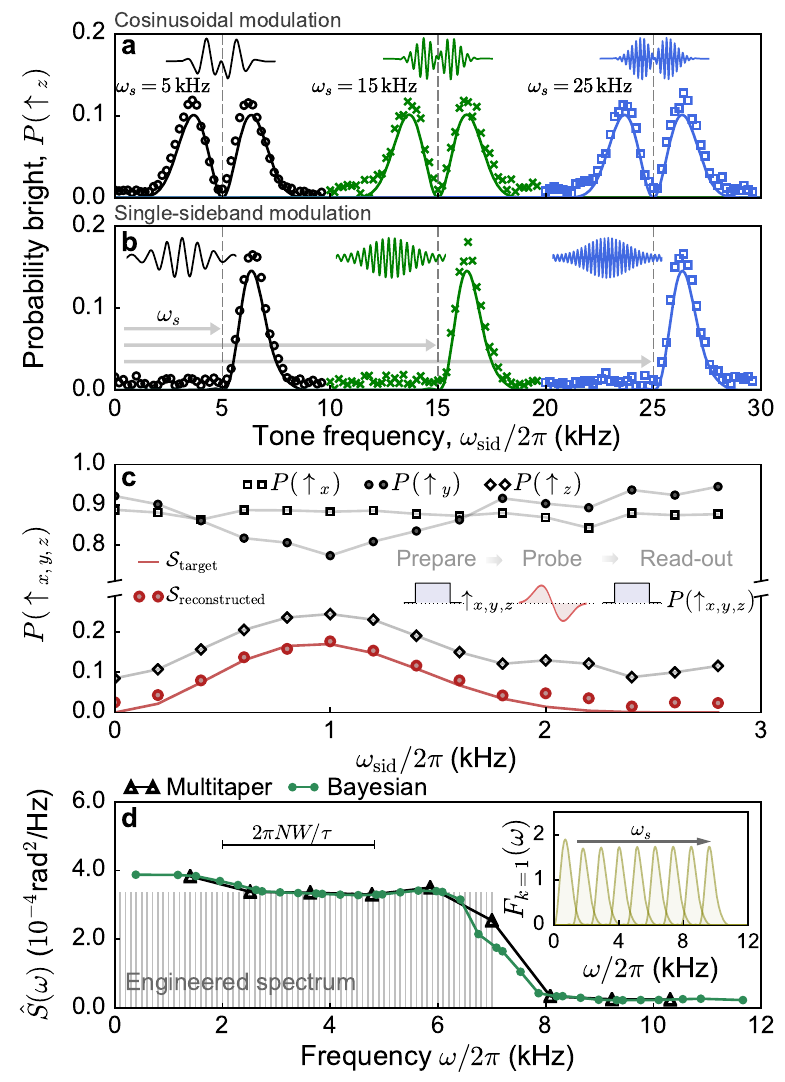}
\caption{Sensing techniques and spectrum reconstruction with DPSS-modulated pulses. \textbf{a)} Measurement and filter function prediction for spectral response of band-shifted first-order DPSS filter functions with cosinusoidal modulation, $\Omega_{\mathrm{mod}, n}^{\text{COS}} 
\equiv v^{(k)}_n (N,W) \cos (n \omega_s \Delta t)$. Modulation depth for system identification measurements was $\alpha=0.5$. The pulse shapes are shown as insets for each modulation frequency. \textbf{b)} Band-shifted single-sideband modulated DPSS, $\Omega_{\mathrm{mod},n}^{\text{SSB}} \equiv v^{(k)}_n (N,W)\cos(n \omega_s \Delta t) \pm \mathcal{H}[ v^{(k)}_n(N,W) ] 
\sin(n \omega_s \Delta t)$, using the same modulation frequencies as in panel \textbf{a}.  $\mathcal{H}$ is the discrete Hilbert transform of the selected DPSS, which is calculated and added to the cosine term in the numeric definition of the modulation envelope before being output from the control system. \textbf{c)} Protocol for three-axes measurements to reconstruct the amplitude filter function in the presence of white dephasing noise with a root-mean-square amplitude of $5 \Omega_x$, where $\Omega_x$ is the maximum Rabi frequency of the control.  Rotations around $x$ and $y$ axes allow preparation and readout along all three axes of the Bloch sphere.  Projective measurements onto $x$, $y$ and $z$ (black markers and lines), expected fidelity under amplitude noise only (red line), and fidelity reconstructed from the three projective measurements (red markers). System identification measurements were all taken with $\alpha=0.75$. \textbf{d)} Reconstruction of an applied noise spectrum (shaded comb teeth) using two approaches (see text).  Inset shows all shifted $F_{\Omega}(\omega)$ for $k=1$.  Here $\omega_s$ ranges from ~$1.4$ to $10.0$\,kHz in steps of ~$1.1$\,kHz.  Horizontal scale bar indicates target bandwidth of each measurement.}
\label{fig3}
\end{figure}

In order to implement DPSS modulated pulses for spectral-reconstruction applications, we apply additional analogue modulation techniques designed to shift the band center from zero to a user-defined frequency $\omega_s$ (Fig.~\ref{fig3}a-b)~\cite{pozar2014omicrowave}. We demonstrate two modulation protocols; cosinusoidal (COS) modulation shifts both positive and negative frequency components by $\omega_s$, 
while single-sideband (SSB)
modulation shifts the band center by $\omega_s$ and suppresses either the positive or negative frequencies, thereby reducing the bandwidth by one half.  Experiments using both techniques demonstrate maintenance of the critical spectral concentration of the DPSS within the shifted bands.  Further details on the implementation appear in the \emph{Supplementary Materials}.

Quantum sensing applications also require the ability to disambiguate changes in the measured operational fidelity due to target signals within a single quadrature, 
\emph{e.g.} $\Omega \sigma_x$, 
from alternate ``interfering'' sources which may be manifested similarly in projective measurements.  For instance, the presence of a Hamiltonian term $\propto{\sigma}_{z}$ during a driven operation $\propto{\sigma}_{x}$ will reduce the measured fidelity of the driven operation in a manner similar to the presence of noise proportional to the control \cite{GreenNJP2013}. Consequently, a single measurement is insufficient to determine which process is at play.  To detect and compensate for such effects, we use tomographic reconstruction~\cite{Poyatos1997Complete, nielsen2002quantum}, by preparing the qubit state along the three Cartesian directions and performing independent sequential measurements in the corresponding bases (Fig.~\ref{fig3}c).  In our experiments we simultaneously apply a target signal $\propto{\sigma}_{x}$ as above, and an additional white dephasing term $\propto{\sigma}_{z}$, which contributes to the sensor's overall response in a way that obfuscates the measurement of the target.  We then isolate the target signal's contribution by combining three projective measurements as ${\cal S}  \equiv (1 + P(\uparrow_x) - P(\uparrow_y) - P(\uparrow_z))/2$. Reconstructed values of ${\cal S}$
(red markers, Fig.~\ref{fig3}c) reproduce the results expected without any ${\sigma}_{z}$-terms well (red line), 
successfully correcting for a vertical offset that would otherwise bias an spectral estimate.

With demonstrations of the relevant band-limited properties of DPSS-modulated qubits, as well as essential band-shifting techniques complete, we move on to demonstrate our spectral reconstruction capabilities.  As a sample application, we reconstruct an engineered amplitude-noise spectrum (Fig. \ref{fig3}d). We employ four different DPSS with $k=1,3,5,7$ and $NW=7$, band-shifted by SSB at nine different modulation frequencies. The spacing of the modulation frequencies was chosen to be about $1/2$ of the bandwidth of the filter functions, which yields measurements with sensitivity in overlapping bands. The various estimates arising from these measurements are combined to produce an reconstruction of the target noise spectrum, which we accomplish this using two distinct techniques. While both approaches are inspired by Thomson's multitaper approach 
\cite{thompson_multitaper}, these also differ in important ways. 

In its original form, multitaper spectral reconstruction aims
to estimate the spectrum of a stationary Gaussian process from a finite set of discrete-time samples.
In this technique, a set of DPSS are used to window the time-domain data in post-processing, producing a set of ``eigenestimates''
of the spectrum in the target band. While each eigenestimate is, in itself, an estimate of the spectrum $\chi^2$-distributed 
about the true value, the key idea of multitaper estimation is to combine different eigenestimates into a weighted sum, leveraging the orthogonality of the DPSS.  This procedure results in a $\chi^2$-distribution with a greater number of degrees of freedom, ensuring consistency and 
increasing variance efficiency ~\cite{thompson_multitaper,percival1993spectral}. 
In order to offset the introduction of out-of-band bias from higher-order DPSS, 
the final estimate is determined through an adaptive weighting procedure, designed to 
favor the lowest-order eigenestimates with the best spectral concentration in the band.

The first reconstruction technique we employ is closest in spirit to the original multitaper, with one crucial 
distinction; by applying DPSS amplitude modulation to the quantum sensor we are, in effect, windowing the noise process {\em before} any measurements are made, and we obtain various eigenestimates via application of different DPSS controls.   This stands in contrast to the manner in which classical multitaper eigenestimates are determined by {\em post}-processing a set of discrete-time samples.  Measured fidelities determine preliminary spectral estimates at the center of each band, which are then weighted according to Thomson's adaptive procedure to obtain a final set of estimates.  

\begin{figure}[t!]
\includegraphics[scale=1]{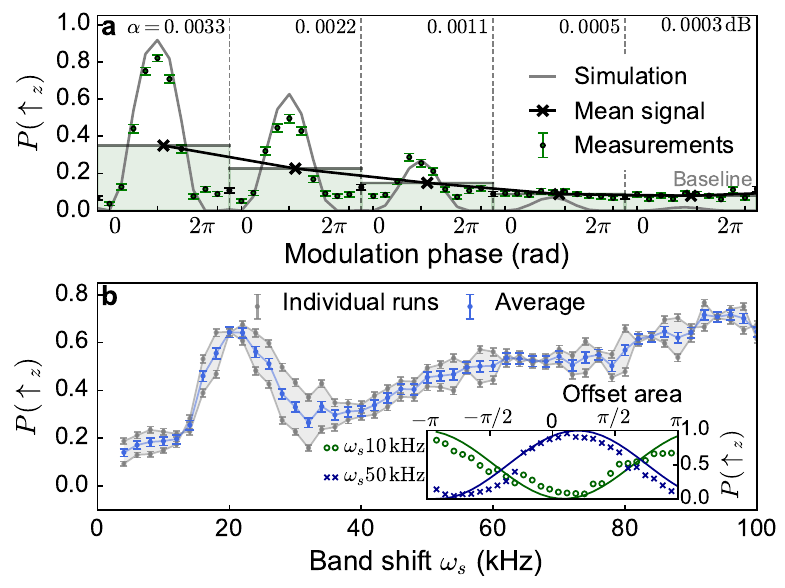}
\caption{Sensing experiments of intrinsic amplitude noise using a single ion. \textbf{a)} Sensitivity calibration using $k=0$ and $\omega_{\mathrm{sid}} = 
\omega_s$, with $\omega_s = \unit[10]{kHz}$. Measurements are taken for linearly sampled phases from $0$ to $2\pi$ with varying modulation depth and interleaved noiseless pulses to identify the fidelity limits of the experiments (``Baseline''). Each measurement is repeated 200 times. The overall signal is represented as the average over all phase realisations. \textbf{b)} Frequency-dependent amplitude instability probed with band-shifted $k=0$ DPSS pulses. The measurement is repeated twice by scanning the $\omega_{s}$ first from $\unit[2-100]{kHz}$ and then from $\unit[100-2]{kHz}$. The bandwidth of the DPSS filter (FWHM) is $\sim\unit[150]{Hz}$. Inset) A small offset pulse is added at the beginning of the main protocol and its area is varied over the range $[-\pi, \pi]$ to demonstrate the frequency-dependent pulse offset behaviour.  High-contrast fringes indicates quantum coherence is maintained.}
\label{fig4}
\end{figure}

In our second approach, each band is sub-divided into a set of smaller segments. For each band, solving a linear inversion determines the Bayesian maximum {\em a posteriori} estimate of the spectrum in each segment, which serves as a preliminary estimate.
Because each segment is contained in multiple bands, the preliminary estimates are then weighted by their Fisher information to determine a final estimate of the spectrum in each segment.   While this approach relies on linear inversion and is thus computationally less efficient than the multitaper technique, the flexibility in the choice of the spectrum model to be used as a prior, as well as the in-band segmentation, allows for improved resolution relative to multitaper in principle. The Bayesian reconstruction in Fig.~\ref{fig3}d uses the multitaper as a prior, and offers slightly improved resolution of the high-frequency cutoff.
Additional technical details relevant to both reconstruction methods are given in the {\em Supplementary Material}. Both procedures produce spectral estimates which quantitatively match the applied spectrum (within resolution limits), and accurately identify the presence of a high-frequency cutoff in the noise.  

We conclude by using DPSS-modulation to determine frequency-resolved information about native noise and nonlinearities in our control system at the end of the synthesis chain (which includes the vector signal generator, an amplifier, cabling, and a waveguide-to-coax converter). For this experiment, we use a single ion and perform DPSS-modulated pulses with $k=0$ producing a net rotation equivalent to $\sim400\;\pi$ rotations, ideally enacting ${\mathbb{I}}$. We calibrate sensitivity to noise by first applying a single-frequency modulation at $\omega_{\mathrm{sid}} = \omega_s$, and averaging over phase (``x'' markers, Fig.~\ref{fig4}a).  We compare this value against interleaved measurements conducted without applied noise to determine the minimum sensitivity achieving SNR $\sim1$.  These measurements demonstrate our ability to detect band-limited amplitude noise with modulation depth as low as $\sim\unit[0.001]{dB}$. 

In measurements taken at different values of $\omega_{s}$, we observe a deviation from the ideal operation over much of the scan range, with a distinct feature around $\unit[20]{kHz}$. We confirm that the measured signals are a manifestation of a frequency-dependent response in the synthesis chain rather than, \emph{e.g.}, extrinsic decoherence, by adding a small phase-shifting pulse at the beginning of the protocol to a (shifted) phase fringe with high contrast (inset, Fig~\ref{fig4}b). 
Investigations into the source of this behaviour are ongoing at the time this manuscript is prepared, although we note that the feature near $\unit[20]{kHz}$ approximately coincides with the maximum Rabi frequency used in this experiment.  Ultimately, this approach provides information that we believe is otherwise inaccessible via independent characterisation of hardware components in our system. 

The demonstrations presented here indicate that appropriately crafted quantum control protocols for qubit-based sensors 
have the ability to overcome significant technical limitations in contemporary quantum sensing experiments.  These protocols can be applied to any nanoscale qubit sensor in which arbitrary phase and amplitude modulation of the driving field is possible and spectral concentration is desired.  It is noteworthy that by reducing the need to account for high-harmonics in the Fourier response of the modulation pattern, the relatively simple and computationally efficient multitaper approach to spectrum reconstruction performs similarly to the more complex Bayesian estimation procedure under the conditions we tested. Future studies will involve the extension of 
DPSS-modulated control to sensing of additive dephasing noise, in order to provide an expanded toolkit of 
band-limited controls for quantum sensors. 


\emph{Acknowledgements}: The authors acknowledge C. Edmunds and A. Milne for contributions to the experimental system and C. Granade and C. Ferrie for useful discussions on Bayesian estimation.  Work partially supported by the ARC Centre of Excellence for Engineered Quantum Systems CE110001013, the US Army Research Office under Contract W911NF-12-R-0012, and a private grant from H. \& A. Harley.

\bibliography{references}
\bibliographystyle{ieeetr}

\end{document}